\documentclass[sigconf, 10pt, pbalance=true]{acmart}

\AtBeginDocument{%
  }

\copyrightyear{2026}
\acmYear{2026}
\setcopyright{cc}
\setcctype{by-nc-nd}
\acmConference[MobiCom '26]{The 32nd Annual International Conference on Mobile Computing and Networking}{October 26--30, 2026}{Austin, TX, USA}
\acmBooktitle{The 32nd Annual International Conference on Mobile Computing and Networking (MobiCom '26), October 26--30, 2026, Austin, TX, USA}
\acmDOI{10.1145/3795866.3844780}
\acmISBN{979-8-4007-2505-0/2026/10}

\begin{document}

\title{WiP: Characterizing and Defending Against Mobile-Agent-Driven MFA Automation}

\author{Yimeng Liu}
\affiliation{%
	Department of Computer Science and Engineering
  \institution{University of California, Merced}\city{Merced, CA}\country{USA}}
\email{yliu327@ucmerced.edu}

\author{Hua Huang}
\affiliation{Department of Computer Science and Engineering
 \institution{University of California, Merced}\city{Merced, CA}\country{USA}}
\email{hhuang80@ucmerced.edu}

\begin{abstract}
Mobile agents automate smartphone tasks by interpreting interfaces, interacting with apps, and coordinating cross-app workflows. This capability challenges the human-mediated separation assumed by passcode-based MFA, creating \emph{factor collapse}: valid authentication factors are combined within one autonomous environment. Our modular pipeline completes all 10 authorized MFA workflows, compared with 3/10 and 6/10 for two single-agent baselines. We also develop a motion-based Android risk signal that distinguishes human- from agent-driven logins with 98.4\% table-top and 92.6\% hand-held accuracy across 116 pilot sessions. These results demonstrate factor collapse and motivate physical-interaction sensing as complementary evidence of user presence.

\end{abstract}



\maketitle

\vspace{-0.05in}
\section{Introduction}




Passcode-based multi-factor authentication (MFA) protects accounts by requiring users to enter a password and then manually transfer a short-lived one-time passcode (OTP) delivered through email or generated by an authenticator application. This workflow implicitly assumes that credential entry, passcode retrieval, and passcode submission remain operationally separated through user participation~\cite{nist80063b4}.

However, this assumption is increasingly challenged by the emergence of
LLM-based mobile agents. Recent LLM-based mobile agents can autonomously operate Android applications across multiple apps and maintain long-horizon execution~\cite{rawles2025androidworld,wu2025assistantsadversaries}. Consequently, an agent possessing valid credentials may retrieve an email-delivered OTP or authenticator code and complete the entire MFA workflow without human participation.
We call this threat factor collapse: an autonomous execution environment joins authentication operations that an MFA deployment assumes are operationally independent. Rather than bypassing MFA, the agent legitimately obtains and submits every required factor within a single automated workflow. In this paper, we study passcode-based MFA, including both email-delivered OTPs and authenticator-generated dynamic codes.

We build a prototype mobile agent that automates complete passcode-based MFA workflows on authorized accounts. Across ten websites and mobile applications, the prototype successfully completes both email-delivered OTP flows and authenticator-app code flows, demonstrating that factor collapse is practical in today's mobile ecosystem.
We further explore whether mobile devices can recover evidence of human participation during passcode entry. Our preliminary results suggest that short-window motion sensing provides a useful complementary signal for distinguishing user-driven from fully agent-driven login sessions.
This paper makes three contributions:



\begin{itemize}
    \item We formulate factor collapse, an emerging threat in which mobile agents operationally collapse passcode-based MFA.

    \item We demonstrate the feasibility of end-to-end MFA automation across ten representative websites and mobile applications.

    \item We present a preliminary device-side sensing defense that detects the absence of ordinary human interaction during authentication.
\end{itemize}

\section{Attack Pipeline and Sensing Defense}
\label{sec:design}

\subsection{Agent-Driven Factor-Collapse Pipeline}

\noindent\textbf{Threat Model.}
We consider a general-purpose mobile agent operating on an unlocked device with GUI-control privileges and the user’s primary credentials. The email or is already configured on the device. The agent must retrieve and submit the code through normal UI actions; we assume no OS compromise, access to authenticator secrets, or lock bypass. This threat arises when a privileged agent is compromised, misdirected, or acts beyond the user’s intended authorization.

\noindent
\textbf{Single-Agent Baselines.}
We evaluate two single-agent baselines across ten services: a basic agent given only the login objective and a detailed agent using platform-specific instructions. Both remain unreliable, failing because of stale passcodes, missed inbox refreshes, premature termination, hallucinated completion, and repeated actions. These results, detailed in the evaluation, show that prompt engineering alone is insufficient for scalable factor-collapse analysis.

\begin{figure}[t]
  \centering
  \includegraphics[width=\columnwidth]{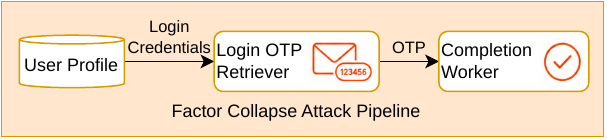}
  \caption{Modular Agent Pipeline for Factor Collapse}\vspace{-0.1in}
  \Description{A high-level workflow in which a mobile agent interacts with a login interface, accesses an authorized passcode channel, completes the login flow, and records the outcome.}
  \label{fig:mfa-auto-outline}
\end{figure}

\noindent
\textbf{Modular Factor-Collapse Pipeline.}
We therefore decompose the MFA workflow into two specialized workers, as
shown in Figure~\ref{fig:mfa-auto-outline}. The \emph{Login OTP
Retriever} receives a controlled account profile, navigates to the target
service, enters the credentials, and detects whether an additional
authentication challenge is presented. When required, it triggers the
challenge, opens the corresponding email or authenticator application,
and extracts the current passcode.

The \emph{Completion Worker} restores the target login context, enters
the retrieved code, completes any remaining confirmation steps, and
determines whether authentication succeeded. The two workers exchange
only the state needed to continue the workflow, including the target
service, account identifier, workflow status, and retrieved passcode.

This decomposition limits each worker's action space, reduces prompt and
context complexity, and isolates failures across application boundaries.
It also enables stage-level recovery: passcode retrieval can be repeated
without re-entering the credentials, and a passcode-entry failure can be
corrected without reprocessing the authentication channel. The resulting
pipeline provides a systematic and reproducible basis for evaluating
mobile-agent-driven factor collapse.

\subsection{Physical-Interaction Sensing Defense}

Factor collapse removes the human participation normally involved in
retrieving and submitting a passcode. We therefore investigate whether
device motion can provide an independent signal of physical user
participation. Prior work shows that mobile motion sensors can capture
behavioral interaction patterns~\cite{giuffrida2014sensed,
fereidooni2023authentisense}.

Our prototype records accelerometer and gyroscope signals
at 50 Hz during the time window covering passcode entry and
final submission. From each window, we compute lightweight
motion features, including signal magnitude, variance,
motion energy, jerk, peak count, and frequency-domain
energy, and classify the session using an extra-trees classifier.
The prediction is used as a risk signal: uncertain or agent-like sessions trigger additional verification rather than being accepted solely on the passcode. This
approach seeks to restore evidence of human participation without
modifying the underlying MFA protocol.

\section{Evaluation}


\begin{figure*}[t]
  \centering
  \begin{minipage}[t]{0.32\textwidth}
    \centering
    \includegraphics[width=\linewidth]{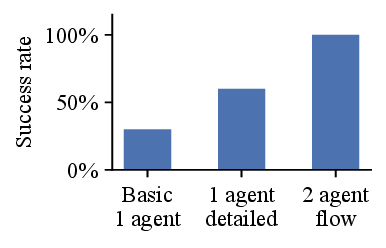}\vspace{-0.1in}
    \caption{Success Rate}
    \Description{A plot summarizing agent success rates for the evaluated MFA automation workflows.}
    \label{fig:agent_success}
  \end{minipage}
  \hfill
  \begin{minipage}[t]{0.32\textwidth}
    \centering
    \includegraphics[width=\linewidth]{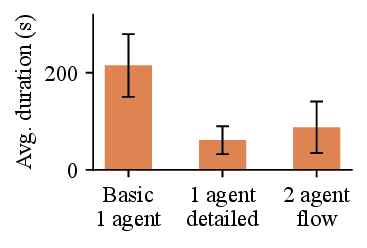}\vspace{-0.1in}
    \caption{Completion Time}
    \Description{Factor-Collapse attack duration for each evaluated MFA automation workflow.}
    \label{fig:agent_duration}
  \end{minipage}
  \hfill
  \begin{minipage}[t]{0.32\textwidth}
    \centering
    \includegraphics[width=\linewidth]{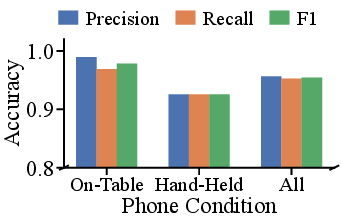}\vspace{-0.1in}
    \caption{Defense Performance }
    \Description{A plot summarizing classification results for the preliminary physical-interaction sensing defense.}
    \label{fig:classification}
  \end{minipage}\vspace{-0.1in}
\end{figure*}

\subsection{Factor-Collapse Implementation}
We implement the pipeline on MobileRun and use GPT-5 mini for its cost–reliability balance in preliminary model tests. The two workers maintain separate prompts and histories, while a lightweight controller shares minimal workflow state and retries failed stages.





\subsection{Factor-Collapse Attack Evaluation}
To evaluate factor-collapse feasibility, we test ten commercial websites and mobile apps spanning multiple service categories. Each account uses either an email-delivered OTP or an authenticator-generated code. We provide the agents with account credentials and instruct them to complete the workflow. We manually verify each outcome and record end-to-end completion time. A trial is counted as successful if the agent reaches the target service's post-login state within a budget of 15 actions; runs that exceed the budget or terminate early are counted as failures. The same budget applies to all three configurations.




Figures~\ref{fig:agent_success} and~\ref{fig:agent_duration} compare the success rate and average completion time, respectively, across the three agent configurations. The basic single-agent design succeeds in only 3/10 of trials and takes more than 200~s on average. Failures commonly result from difficulty locating less salient interface elements, refreshing the mailbox for newly delivered OTPs, and identifying the relevant message in a cluttered inbox. These inefficiencies often exhaust the same step budget before login is completed.

Providing detailed instructions increases the success rate to 6/10 and substantially reduces completion time by replacing exploratory actions with explicit workflow guidance. The two-agent design achieves a 10/10 success rate in our experiments. Assigning form interaction and OTP retrieval to separate agents reduces context switching and limits interference from irrelevant inbox content. Although coordination modestly increases completion time relative to the detailed single-agent design, it substantially improves reliability.

\subsection{Physical-Interaction Sensing Evaluation} We implement a preliminary Android defense prototype with a standard MFA login interface for entering account credentials and requesting an OTP. The prototype also includes a lightweight physical-interaction sensing module that records accelerometer and gyroscope data at 50 Hz throughout the login session. It then applies the detection algorithm described in the Section 2.2 to classify each session as user-driven or agent-driven.

We evaluate the prototype under two common device-use conditions: the phone resting on a table and the phone held in hand. In each condition, login sessions are completed either manually by a participant or automatically by our factor-collapse agent. We recruited four participants and collected 116 sessions in total: 62 table-top sessions (31 user-driven and 31 agent-driven) and 54 hand-held sessions (27 user-driven and 27 agent-driven). We evaluate the classifier using five-fold cross-validation, with each fold maintaining a balanced distribution of user-driven and agent-driven sessions. Performance is reported using accuracy, precision, recall, and F1 score.



Figure~\ref{fig:classification}  reports classification performance under the table-top and hand-held conditions, as well as across all sessions. In the table-top condition, user-driven and agent-driven sessions are nearly perfectly separable, with the classifier achieving 98.4\% accuracy, with 0.989 precision, 0.969 recall, and an F1 score of 0.978. When the phone rests on a surface, agent operation produces little device motion, whereas manual interaction introduces touch-related disturbances. The hand-held condition is more challenging because both user-driven and agent-driven sessions contain natural hand motion. Nevertheless, the classifier achieves 92.6\% accuracy and an F1 score of 0.926 across 54 sessions. When data from both conditions are combined, the classifier achieves 95.7\% accuracy, with 0.956 precision, and 0.953 recall and 0.954 F1. Overall, these results suggest that user-driven and agent-driven logins exhibit distinguishable physical-interaction patterns.



\section{Conclusion and Future Work}


We identify \emph{factor collapse}: a general-purpose mobile
agent can combine credential entry, passcode retrieval, and
passcode submission within one autonomous workflow. On
ten selected targets, our modular two-agent design completes
10/10 workflows, compared with 3/10 and 6/10 for two
single-agent baselines. A pilot motion-sensing classifier also
distinguishes user-driven from agent-driven sessions,
including 92.6\% accuracy under hand-held use. 

Future work will formalize the threat model and expand the evaluation across additional services, devices, participants, and authentication workflows. We will also examine adaptive agents, sensor manipulation, and deployment conditions that may reduce the reliability of physical-interaction sensing. Finally, we will investigate how sensing-based risk signals can be combined with explicit user confirmation, user-verifying authenticators, and verifier-bound authentication to preserve user intent without relying solely on the successful submission of a passcode.

\begin{acks}
This work is partially supported by the U.S. National Science Foundation under Grant No.~2543097.
\end{acks}

\vspace{-0.1in}
\bibliographystyle{ACM-Reference-Format}
\bibliography{bib}

@techreport{nist80063b4,
  author      = {Temoshok, D. and others},
  title       = {Digital Identity Guidelines: Authentication and Authenticator Management},
  institution = {NIST},
  number      = {SP 800-63B-4},
  year        = {2025},
  doi         = {10.6028/NIST.SP.800-63B-4}
}

@misc{rawles2025androidworld,
  author        = {Rawles, C. and others},
  title         = {{AndroidWorld}: A Dynamic Benchmarking Environment for Autonomous Agents},
  year          = {2025},
  eprint        = {2405.14573},
  archivePrefix = {arXiv},
  doi           = {10.48550/arXiv.2405.14573}
}

@misc{wu2025assistantsadversaries,
  author        = {Wu, L. and others},
  title         = {From Assistants to Adversaries: Exploring the Security Risks of Mobile {LLM} Agents},
  year          = {2025},
  doi           = {10.48550/arXiv.2505.12981}
}

@misc{giuffrida2014sensed,
  author       = {Giuffrida, C. and others},
  title        = {I Sensed It Was You: Authenticating Mobile Users with Sensor-Enhanced Keystroke Dynamics},
  howpublished = {DIMVA},
  year         = {2014},
  doi          = {10.1007/978-3-319-08509-8_6}
}

@misc{fereidooni2023authentisense,
  author       = {Fereidooni, H. and others},
  title        = {{AuthentiSense}: A Scalable Behavioral Biometrics Authentication Scheme Using Few-Shot Learning for Mobile Platforms},
  howpublished = {NDSS},
  year         = {2023},
  doi          = {10.14722/ndss.2023.23194}
}

\end{document}